# Enhancing Perovskite Electrocatalysis through Strain Tuning of the Oxygen Deficiency

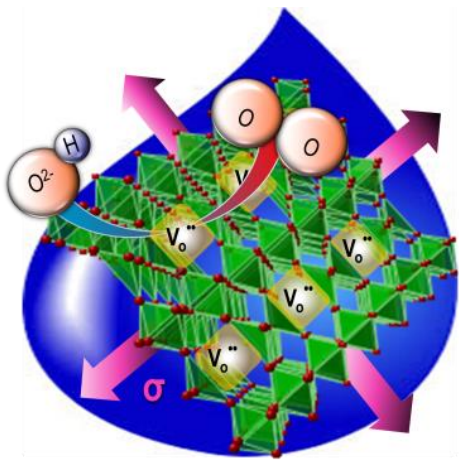

Jonathan R. Petrie[+], Hyoungjeen Jeen[+], Sara C. Barron[±], Tricia L. Meyer[+], and Ho Nyung Lee[+*]

[+]*Materials Science and Technology Division, Oak Ridge National Laboratory, Oak Ridge, TN, 37831, USA.*

[±]*National Institute of Standards and Technology, 100 Bureau Dr, Gaithersburg, Maryland 20899, USA.*

*Supporting Information Placeholder*

**ABSTRACT:** Oxygen vacancies in transition metal oxides facilitate catalysis critical for energy storage and generation. However, it has proven elusive to promote vacancies at the lower temperatures required for operation in devices such as metal-air batteries and portable fuel cells. Here, we use thin films of the perovskite-based strontium cobaltite ($SrCoO_x$) to show that epitaxial strain is a powerful tool towards manipulating the oxygen content under conditions consistent with the oxygen evolution reaction, yielding increasingly oxygen deficient states in an environment where the cobaltite would normally be fully oxidized. The additional oxygen vacancies created through tensile strain enhance the cobaltite's catalytic activity towards this important reaction by over an order of magnitude, equaling that of precious metal catalysts, including $IrO_2$. Our findings demonstrate that strain in these oxides can dictate oxygen stoichiometry independent of ambient conditions, allowing unprecedented control over oxygen vacancies essential in catalysis near room temperature.

Incremental changes in oxygen vacancies have leveraged large shifts in the electrocatalytic properties of transition metal oxides (TMOs).[1] Due to the catalytic effect of these oxygen deficiencies, changes in oxygen stoichiometry within binary oxides, such as $TiO_2$, are known to affect water-splitting and the oxygen evolution reaction (OER), which are important for energy production and storage devices, including photocatalysis, metal-air batteries, and fuel cells.[2,3] In addition to binary oxides, studies on more complex oxides, such as Kim, et al. on $CaMnO_{2.5}$,[4] have also suggested that oxygen vacancies near the surface can affect catalysis due to an increase in the number of active sites around these defects,[5] a weaker metal-oxygen bond yielding a faster intermediate exchange,[6] and vacancy-induced electron-doping that changes the spin configuration to $e_g^1$ for more efficient electron transfer.[7] Moreover, by controlling the oxygen anion, rather than metal cation, concentration, the impurity and defect segregation complexities associated with heterovalent cation doping is minimized.[6a, 8]

However, the highly-oxidizing conditions associated with the OER limits the ability to retain oxygen deficiencies for such electrocatalysis. To truly functionalize the catalytic potential of oxygen vacancies in TMOs, a new parameter is required to allow control over the oxygen stoichiometry in surroundings that would normally suppress oxygen deficiencies. Recently, we have used epitaxial thin films of strontium cobaltite, $SrCoO_x$ (SCO) to demonstrate a relationship between strain and oxygen content in aprotic annealing environments of several hundred degrees Celsius.[9] This oxide has a topotactic transition between the brownmillerite phase $SrCoO_{2.5}$ (BM-SCO) and the oxidized perovskite phase $SrCoO_{3-\delta}$ (P-SCO), where $0 \leq \delta \leq 0.25$.[9b, 9c] Due to the easy intercalation of $O^{2-}$ through BM-SCO offered by its open framework and metastability of $Co^{4+}$ in P-SCO, it has an exceptionally low oxygen activation energy (<1 eV) that changes by tenths of an eV under strain.[10] Due to this intriguing blend of facile oxygen incorporation and oxygen activity, SCO was seen as an ideal candidate for examining the catalytic possibilities of strain-induced oxygen deficiencies in TMOs.

In this work, we use epitaxial strain engineering to systematically tune the oxygen stoichiometry of P-SCO thin films in conditions consistent with the critically-important OER.[11] Without such strain, the highly oxidizing environment during OER results in fully oxidized P-SCO with few vacancies.[12] After validating a strain-induced change in oxygen stoichiometry both in the bulk and near the surface, we have then used this control of oxygen content to artificially augment the oxygen vacancies in P-SCO and yield significantly enhanced OER activities. These activities compare favorably to a thin film of $IrO_2$ tested under the same conditions, demonstrating the promise of this method.

To monitor the topotactic oxidation to P-SCO in OER conditions, a set of pre-oxidized BM-SCO films were epitaxially grown on lattice-mismatched substrates using pulsed laser epitaxy (PLE). These films had uniform film thicknesses of 15 nm to ensure minimal strain relaxation on various perovskite substrates. The substrates included (001) $(LaAlO_3)_{0.3}$-$(SrAl_{0.5}Ta_{0.5}O_3)_{0.7}$ (LSAT), (001) $SrTiO_3$ (STO), (110) $DyScO_3$ (DSO), (110) $GdScO_3$ (GSO), and (001) $KTaO_3$ (KTO), whose pseudo-cubic parameters varied, respectively, from $a_{sub}$ = 3.868 to 3.989 Å. While BM-SCO is orthorhombic ($a_o$ = 5.574, $b_o$ = 5.447, $c_o$ = 15.745), stoichiometric P-SCO is cubic with $a_c$ = 3.829 Å, leading to substrate-induced tensile strains from ε = 1.0 to 4.2% for fully oxidized SCO. The substrates were attached to glassy carbon (GC) rods. All films included a 10 nm thick $La_{0.8}Sr_{0.2}MnO_3$ (LSMO) conducting underlayer to ensure uniform charge transport from the rod to the cobaltite layer on non-conducting substrates.[13] In addition to its known inactivity to OER,[13] the conducting LSMO allowed us to eliminate the use of a carbon binder used in other thin film studies, which may introduce an additional current under testing due to carbon degradation or hybridization.[14] BM-SCO films were initially topotactically oxidized to P-SCO through an *ex-situ* electrochemical oxidation for 5 minutes at 1.6 V vs RHE in a fully oxygenated 0.1 M KOH solution, which is consistent with OER conditions and has been shown to fully oxidize unstrained SCO.[12, 15] Subsequent potentiodynamic and galvanostatic anodization scans confirmed, respectively, OER activity and a measure of stability (see Experimental Details). XRD reciprocal space mapping substantiated that, after oxidation from BM-SCO, all P-SCO films on the aforementioned substrates were coherently strained in-plane (see Figure S1 and S2), leaving the out-of-plane c-parameter as the only free structural variable.

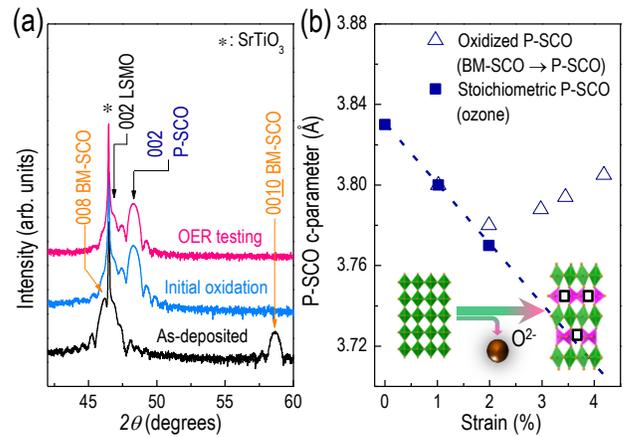

*Figure 1.* a) Representative XRD θ-2θ scans of an as-deposited BM-SCO/LSMO film and an electrochemically oxidized P-SCO/LSMO film on a (001) STO substrate. The scans after the initial oxidation at 1.6 V for 5 minutes and post-OER testing are shown. b) Deviation of the out-of-plane c-parameter from a stoichiometric Poisson ratio~0.26 is due to oxygen vacancy formation. These vacancies can be found in oxygen deficient layers (pink) that alternate with fully oxidized octahedral layers (green).

Figure 1a shows a representative XRD θ-2θ scan around the P-SCO 002 peak for both pre-oxidized BM-SCO and electrochemically oxidized P-SCO on a STO substrate (ε = 2.0%). The initial oxidation for 5 minutes at 1.6 V is displayed as well as the same P-SCO film after OER activity and stability testing. All peaks are clearly defined with Kiessig fringes that verify the superior film quality and provide a measure of thickness. While the LSMO 002 peak shows no change after oxidation, the lack of both BM-SCO 008 and half-order 00$\underline{10}$ peaks indicates a full conversion to P-SCO ($\delta \leq 0.25$) in all films. Along with no change in the X-ray reflectivity (not shown), the similarity between the P-SCO both before (initial oxidation) and after OER testing suggests that amorphization and degradation throughout the entire film is not occurring. In Figure 1b, the P-SCO c-parameter vs strain is compared to previous bulk (unstrained) P-SCO studies or to prior studies on ozone-deposited P-SCO.[9a, 9b, 12, 15] In these studies, the P-SCO is nearly fully stoichiometric and there is a clear linear shift in the c-parameter that can attributed to a Poisson-type contraction with a ratio $\nu \sim 0.26$. However, the monotonic shift in the c-parameter for our coherently deposited, electrochemically oxidized films increasingly diverges from this stoichiometric behavior with tensile strain ε > 1%. Since an increase in the oxygen vacancies is known to result in lattice expansion for perovskite-type oxides, this uniaxial divergence in the only unconstrained direction (c-parameter) is attributed to increasing oxygen deficiencies throughout the films.[16] As the tensile strain on the SCO film intensifies from ε = 1.0 to 4.2%, the increasing difference from the stoichiometric behavior signifies a growth in the amount of oxygen vacancies.

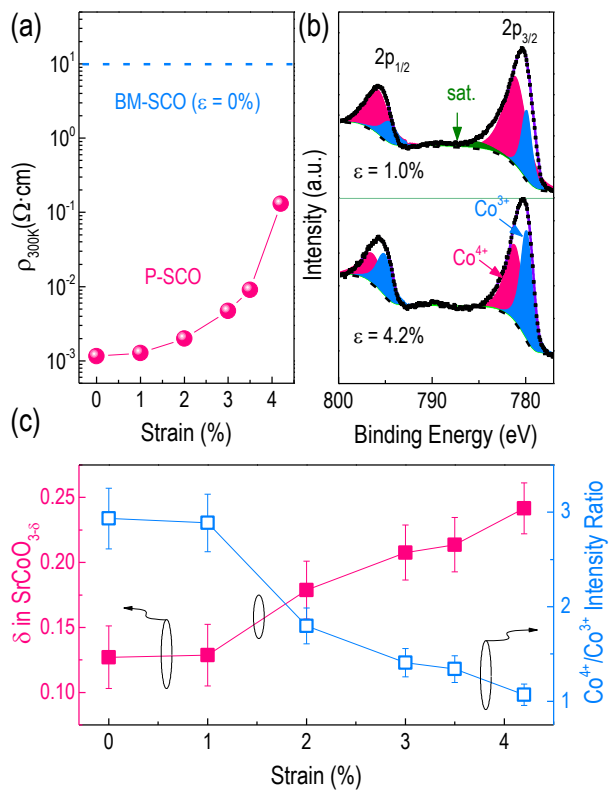

*Figure 2.* a) The increasing resistivity of the strained P-SCO films at room temperature due to strain-induced loss of oxygen. b) The XPS Co-2p and satellite (sat.) peaks of electrochemically oxidized P-SCO on LSAT ($\varepsilon$ = 1.0%) and KTO ($\varepsilon$ = 4.2%) substrates. The fit is in purple. c) By plotting the ratio of the intensity of the Co$^{4+}$ to Co$^{3+}$ peaks, a clear trend towards decreasing Co valency with tensile strain is shown, with a matching trend in oxygen non-stoichiometry ($\delta$). A strain relaxed P-SCO film on LSAO was used for $\varepsilon$ = 0%.

Electrical *dc* transport measurements at room temperature (see Figure 2a) also support this systematic change in stoichiometry for $\varepsilon$ > 1%. A 15 nm thick, strain relaxed P-SCO film on SrLaAlO$_4$ (LSAO) was used as an unstrained control ($\varepsilon$ = 0%). As the tensile strain increases, the film becomes less conducting, suggesting a higher oxygen vacancy concentration.[9c] Indeed, as seen in Figure S3 for the resistivity over a 5 K to 300 K temperature regime, only the $\varepsilon$ ≤ 1% strained film displays metallicity ($\delta$ ~ 0.1) as the $\varepsilon$ = 1% again shows similar properties to the unstrained film; the others are not metallic due to the vacancy-induced disruption of the exchange responsible for such behavior in fully-oxygenated P-SCO.[9b] However, none of the films display the high resistivities typically found in a BM-SCO film of comparable thickness, putting an upper boundary on the oxygen deficiency at $\delta$ ≤ 0.25 for P-SCO.[17]

To further verify the change in oxygen content near the potentially catalytic surface, we investigated the strained P-SCO films via x-ray photoelectron spectroscopy (XPS) using the Co 2p peaks.[18] The oxygen stoichiometry was estimated by charge-compensating oxygen loss with a lower Co oxidation state.[9c] As expected from the XRD and electronic transport data, the unstrained P-SCO displays a spectra similar to the $\varepsilon$ = 1.0% film. As seen for the spectra of the $\varepsilon$ = 1.0 and 4.2% films in Figure 2a, the shoulders at higher energies decrease with a greater degree of tensile strain, indicting a spectral weight transfer from the fully oxidized Co$^{4+}$ to lower energy Co$^{3+}$ peaks. Satellites related with the shake-up structure are also included. The shift in Co valency can be quantified in Figure 2b, where the intensity ratio of the Co$^{4+}$ to Co$^{3+}$ peak falls from ~3 to ~1 between the minimum and maximum tensile strains. Between the $\varepsilon$ = 1.0 to 4.2% films, we can estimate an increase of 0.3e-, which is compatible with $\Delta\delta$ ≤ 0.15 in SrCoO$_{3-\delta}$ as the equilibrium state transitions from SrCoO$_{2.9}$ to SrCoO$_{2.75}$ with tensile strain.

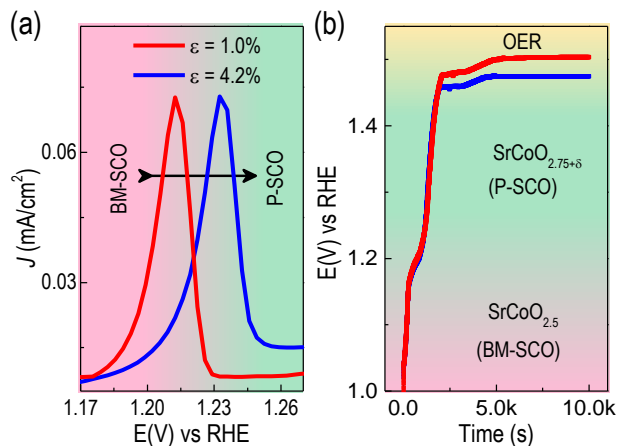

*Figure 3.* a) Anodic scan around the topotactic transformation peak as BM-SCO is oxidized to P-SCO and b) galvanostatic stability scans at 5 µA of SCO/LSMO on LSAT ($\varepsilon$ = 1.0%) and KTO ($\varepsilon$ = 4.2%) at 5 mV/s in O$_2$-saturated 0.1 M KOH.

Electrochemical tests, including *iR*-corrected cyclic voltammagrams (CVs), galvanostatic anodization, and polarization curves, were performed to probe the oxidation of strained SCO and its OER activity as well as its stability under OER conditions. Figure 3a shows representative peaks from CV scans on the $\varepsilon$ = 1.0 and 4.2% films for the oxidative BM-SCO (SrCoO$_{2.5}$) to P-SCO (SrCoO$_{2.75}$) topotactic transformation. Films at intermediate strains display intermediate shifts between these two strain values. There is an anodic shift with tensile strain in both this peak and additional perovskite oxidation (as seen in the full CV in Figure S4). In our previous theoretical work on annealed SCO, DFT calculations revealed that such strain raises the ground state energy of O$^{2-}$ in a potential vacancy site due to a decline in the stabilizing effects of hybridization between the Co 3d and O 2p orbitals with lengthening Co-O bond length.[9a, 19] Here, the anodic peak shift signifies that the driving potential to intercalate oxygen into the film rises with growing tensile strain, which facilitates vacancy generation. Galvanostatic measurements (see Figure 3b) at 5 µA indicate that less strained films are more heavily oxidized and that all films are stable for an extended period of time.[12] To further verify that the films were chemically stable after testing, XPS measurements were used to compare the Sr/Co composition ratio between P-SCO to BM-SCO, as shown in Figure S5. While P-SCO is always enriched in Sr near the surface, there is no statistical difference between the composition of initially oxidized P-SCO and P-SCO that had undergone OER testing.

Polarization curves for the strained P-SCO are shown in Figure 4. Since these oxides are epitaxial films and not in nanoparticulate form within a carbon matrix, it is difficult to directly compare the activities in this study with those thin film results[20]. Therefore, a 50 nm thick (111) $IrO_2$ thin film, representative of a high activity noble metal catalyst, was deposited on GC by physical vapour deposition (PVD) and tested under the same conditions.[21] As the tensile strain in P-SCO is increased from 0 to 4.2%, the onset potential for OER, as defined by the intersection of the tangents from the linear portion of each curve, is reduced by ~100 mV towards the onset of $IrO_2$. The inverse relationship between catalytic activity and conductivity (Figure 2a) as well as thickness studies on P-SCO (Figure S6) suggests that here, as opposed to $LaCoO_3$, charge transfer considerations are not hindering the OER reaction.[13] To ensure that the LSMO is not affecting activity, Figure S5 also compares P-SCO/LSMO on the non-conducting STO substrate used throughout this study to P-SCO on a conducting Nb-doped STO substrate, resulting in minor differences in activity. Furthermore, Tafel slopes (see Figure S7) are ~40 mV/dec for all strained oxides and $IrO_2$ films, suggesting a similar reaction mechanism that is enhanced by oxygen non-stoichiometry at the surface.[21-22] Finally, since CV/galvanostatic oxidation measurements indicate the OER as the dominant reaction at 1.6 V,[6a] we quantitatively compared the activity of these films by observing $J$ at this common OER potential.[12] As represented in Figure 4b, these activities rose by over an order of magnitude for films containing increased oxygen vacancies via application of modest tensile to ~4%, becoming comparable to the activity of $IrO_2$. [21, 23]

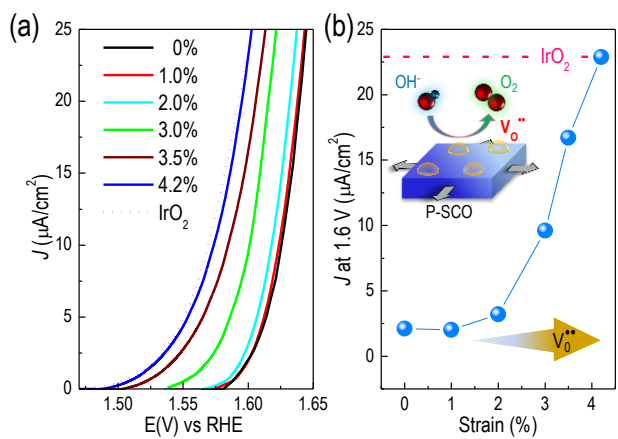

*Figure 4.* Evolution of OER activity with epitaxial stain. a) Polarization curves for the OER reaction on P-SCO under increasing amounts of biaxial tensile strain. b) Current densities at 1.6 V vs RHE for all films are plotted as a function of strain. The activity for a textured (111) $IrO_2$ film is included.

In summary, we establish through epitaxially-straining $SrCoO_{3-\delta}$ thin films that oxygen deficiencies in the cobaltite can be tailored from $\delta \leq 0.1$ to $\delta \sim 0.25$ in aqueous, highly-oxidizing environments. These shifts in the oxygen content occur through modest amounts of tensile strain over 1%, resulting in anion concentrations not possible before in the unstrained bulk material. This newfound ability to control the oxygen vacancy concentration during OER allows the enhancement of the cobaltite's activity by more than an order of magnitude, commensurate with the catalytic performance of more expensive fully-oxidized noble metal electrodes, such as $IrO_2$. Such an increase in activity is likely due to the vacancies themselves or the resultant increase in $Co^{3+}$, which, if in the intermediate spin state, can act as an $e_g^1$ perovskite catalyst first hypothesized by Suntivich, et al [7] Moreover, this increase in activity is not accompanied by a decrease in structural or chemical stability, as found in other oxides reliant on cation, rather than anion, substitution.[6a] Consequently, the realization that strain can decouple the traditional relationship between oxygen and electrochemical operating conditions promises new functionalities in metal oxides for energy applications.

## ASSOCIATED CONTENT

### Supporting Information

Experimental details; XRD θ-2θ scans and reciprocal space maps (RSMS) for BM-SCO and P-SCO; electrical transport from 2K to 300K; CVs of strained films; XPS chemical stability measurements; effects on OER activity from either LSMO underlayer or P-SCO film thickness; Tafel curves. This material is available free of charge via the Internet at http://pubs.acs.org.

## AUTHOR INFORMATION

### Corresponding Author

*Ho Nyung Lee
hnlee@ornl.gov

## ACKNOWLEDGMENT

This work was supported by the Laboratory Directed Research and Development Program of Oak Ridge National Laboratory, managed by UT-Battelle, LLC, for the U. S. Department of Energy (electrochemical characterization) and by the U.S. Department of Energy, Office of Science, Basic Energy Sciences, Materials Science and Engineering Division (synthesis and physical property characterization). Reference electrode depositions were performed as a user project at the Center for Nanophase Materials Sciences, which is a DOE Office of Science User Facility.

Supporting Information

# Enhancing Perovskite Electrocatalysis through Strain Tuning of the Oxygen Deficiency

*Jonathan R. Petrie[+], Hyoungjeen Jeen[+], Sara C. Barron[±], Tricia L. Meyer[+], and Ho Nyung Lee[+*]*

[+]Materials Science and Technology Division, Oak Ridge National Laboratory, Oak Ridge, TN, 37831, USA.

[±]National Institute of Standards and Technology, 100 Bureau Dr, Gaithersburg, Maryland 20899, USA

**EXPERIMENTAL SECTION**

**Thin Film Synthesis:** Epitaxial films of BM-SCO were grown 15-nm thick on different substrates through pulsed laser epitaxy (PLE).[1] The film growth temperature, oxygen partial pressure, laser fluence, and repetition rate were fixed at 750 °C, 100 mTorr, 1.5 J/cm$^2$, and 5 Hz, respectively. Prior to BM-SCO growth, 10-nm thick La$_{0.8}$Sr$_{0.2}$MnO$_3$ (LSMO) underlayers were deposited under the same conditions but at a lower growth temperature (625 °C). The 50-nm thick IrO$_2$ films seen in figure S6 were deposited onto (001) STO substrates via PVD.

**Electrochemical Characterization:** The electrochemical oxidation of BM-SCO was performed in a 150 ml solution of O$_2$-saturated 0.1 M KOH developed with Sigma-Aldrich KOH pellets and Milli-Q water. The oxygen saturation through a bubbler both maintains a constant oxygen concentration as well as minimizing CO$_2$ poisoning from the positive total pressure in the cell compared to the surrounding atmosphere. A three-electrode setup was used with a Pt counter electrode and standard calomel reference electrode (SCE). For conversion to RHE, the reference electrode was calibrated to the reduction of H$_2$ at the counter electrode. Samples were diced into 2.5 x 2.5 mm$^2$ and the lattice mismatched substrates attached via conductive paste (e.g. silver or carbon-based) to a polished glassy carbon (GC) disc (5 mm in diameter). Electrical conductivity was maintained from the GC to the film by applying the conductive paste along the side of the substrate to the deposited films. All paste was subsequently covered with epoxy to prevent any reaction in solution. A metallic LSMO underlayer was deposited below the SCO to ensure that the electrochemical activity is not affected by any differences in conductivity (see Figure S5). Potential was applied via a Biologic SP-200 Potentiostat. Ohmic losses due to the film and solution were determined via a high frequency (~100 kHz) impedance measurement and subtracted from the applied potential to obtain *iR*-corrected currents. The electrochemical characterization was all done in an O$_2$-sat 0.1 M KOH solution to ensure reproducibility. Beginning with the initial BM-SCO film, the steps consisted of:

(1) cycling from 1.0 to 1.65 V at 10 mV/s for a single CV. Representative scans appear in Figure 3a for the films with the lowest and highest lattice mismatch.

(2) a linear sweep to 1.6 V before being held there for 5 minutes to ascertain a steady-state stoichiometry. Films were characterized via XRD for structure. These scans can be seen in Figure 1 and Supplementary Figure 2.

(3) cycling at least 50 times at 50 mV/s between 1.0 and 1.65 V to ensure a stable film surface. No peaks indicative of solution in contact with the LSMO underlayer were observed.

(4) linear voltammetry sweeps from 1.0 to 1.65 V and back at 5 mV/s and a rotating disk speed of 1600 rpm to determine OER steady-state polarization curves. These were repeated at least three times to ensure reproducibility. The average of the anodic and cathodic sweeps were used to minimize capacitive effects. The results can be seen in Figure 4a. Above ~1.5 V vs RHE, all anodic currents were attributed to oxygen evolution as further anodic peaks associated with continued P-SCO oxidation were not established in either $O_2$- or Ar-saturated solutions.[2]

(5) galvanostatic anodic measurements at 5 μA/cm² (near the current density of P-SCO on STO) for 10,000 seconds as an initial screen of stability, as seen in Figure 3b.

Specific surface area measurements after testing were used to minimize the relative effect of different surface areas on current density. Consistent with prior work on benchmarking OER catalysts, the electrochemically-active specific area (ESCA) of these P-SCO films were determined via double-layer capacitance measurements around the open-circuit potential (OCP)[3].

**Structural and Spectroscopic Characterization:** The sample structure was characterized with a high-resolution four circle XRD. Temperature-dependent DC transport measurements were conducted using the van der Pauw geometry with a 14 T Physical Property Measurement System (PPMS) and a scan rate of 3K/s. Valence measurements via XPS were performed *ex-situ* within 24 hours to minimize oxygen loss. The photoelectron spectra were acquired using monochromatic Al Kα X-rays in a Kratos Axis Ultra DLD spectrometer at 0.1 eV steps with a pass energy of 20 eV. The Co *2p* region was energy adjusted to C 1s at 285 eV and fit with a Shirley background; each peak was fit with an 80% Gaussian/20% Lorentzian curve to represent the either the +3 or +4 Co valence state. Based on prior references, $Co^{3+}$ was assigned a *2p3/2* peak ~ 780 eV and $Co^{4+}$ a *2p3/2* peak ~ 781 eV.[4] Satellite peaks for *2p3/2* were taken into account so as not to overestimate the amount of $Co^{4+}$ in the materials system. Systematic errors are added in quadrature.

**FIGURES**

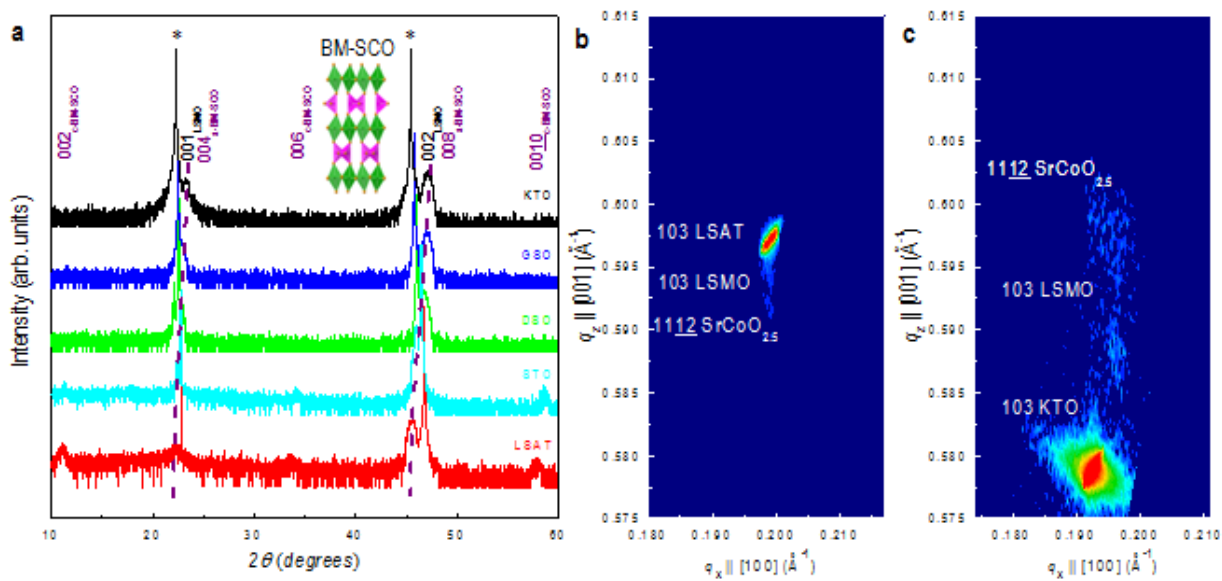

**Figure S1. Coherent growth of BM-SCO (SrCoO₂.₅) /LSMO on all substrates.** (a) XRD $\theta$-$2\theta$ scans of as-deposited BM-SCO/LSMO films on various substrates. Due to the better lattice mismatch along the *c*-axis of BM-SCO, the films on DSO, GSO, and KTO are *a*-axis-oriented, while the rest films on LSAT and STO are *c*-axis oriented. Substrate peaks are noted by asterisks (*) and shifts in (008) and (004) BM-SCO peaks with dashed lines. (b) RSMs of LSAT ($\varepsilon$ = 1.0%) and KTO ($\varepsilon$ = 4.2%) are shown to confirm the coherent growth on the least and most tensile-strained substrates, respectively. Interestingly, although the BM-SCO on KTO appears coherently strained KTO, the LSMO shows some slight contraction.

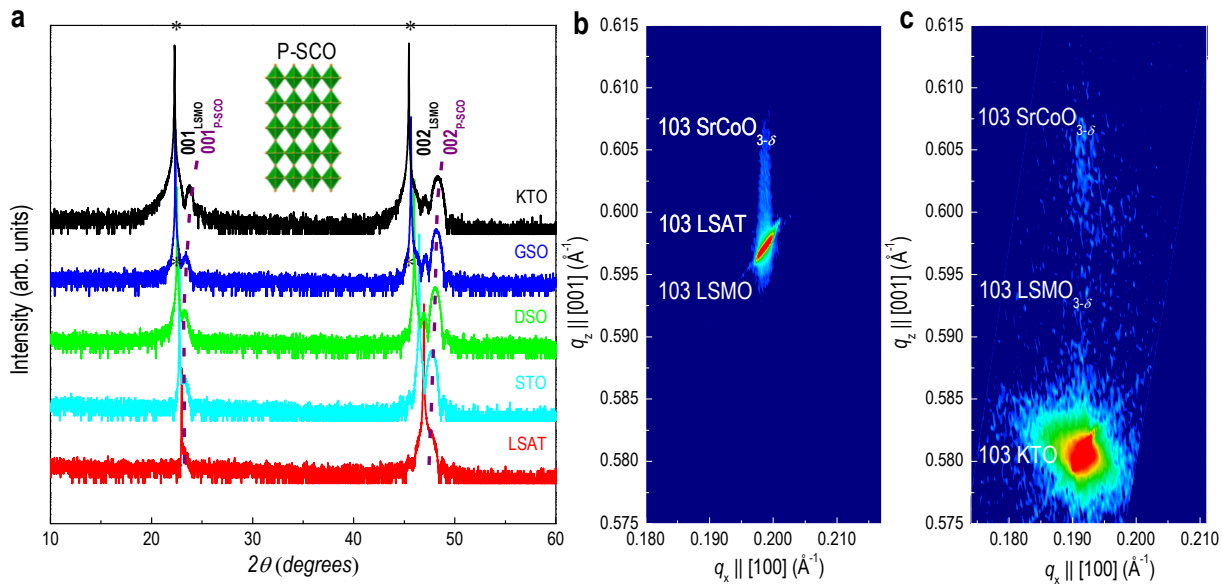

**Figure S2. Coherent (*ex-situ* electrochemical) topotactic oxidation of P-SCO (SrCoO$_{3-\delta}$) /LSMO on all substrates.** (**a**) XRD *θ-2θ* scans of topotactically oxidized P-SCO films on lattice-mismatched substrates. The lack of half-order peaks and the shift in the 002 peak are indicative of the perovskite SrCoO$_{3-\delta}$ (P-SCO), where $\delta \leq 0.25$. Substrate peaks are marked by asterisks and shifts in (002) and (001) P-SCO peaks with dashed lines. (**b**) RSMs of P-SCO films on LSAT and KTO are shown to confirm coherent growth on the least and most tensile-strained substrates, respectively. After oxidation, both P-SCO and LSMO KTO appear coherently strained on KTO.

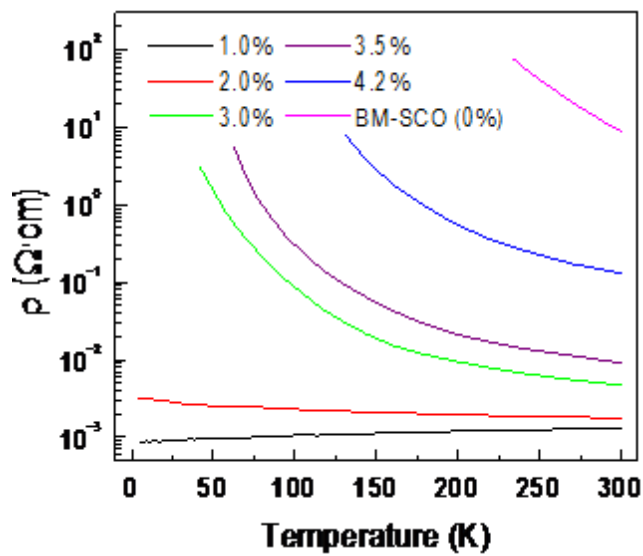

**Figure S3. Change in electrical resistivity for preferential oxygen loss in tensile strained P-SCO.** The resistivity of the electrochemically oxidized films at 300 K indicates the increasing absence of oxygen with tensile strain. As grown BM-SCO on STO is included as a reference, putting an upper boundary on the oxygen deficiency for P-SCO at $\delta \leq 0.25$.

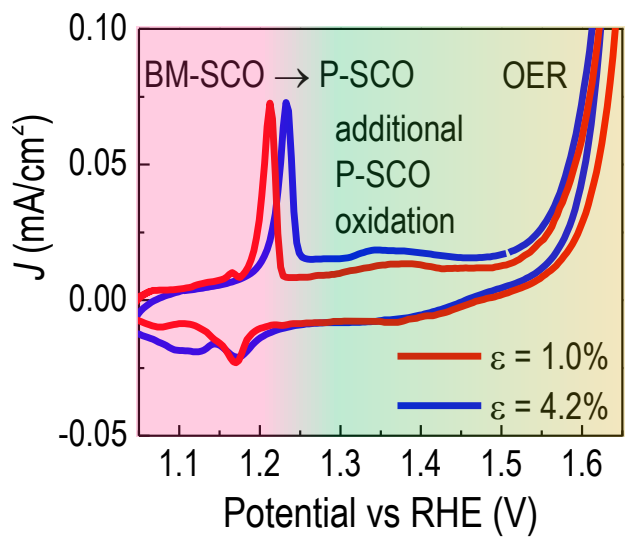

**Figure S4. Changes in cyclic voltammetry in representative tensile strained P-SCO.** Between the $\varepsilon$ = 1 and 4.2% films, there is a readily apparent anodic shift in both the topotactic transformation peak and additional oxidation of the P-SCO with tensile strain at 10 mV/s and $O_2$-sat 0.1 M KOH. One can also note that the OER appears to visibly onset in the more tensile strained film, an observation that was verified in the later polarization scans using RDE. Films with strains intermediate between 1% and 4.2% had corresponding intermediate shifts in peaks and OER onset.

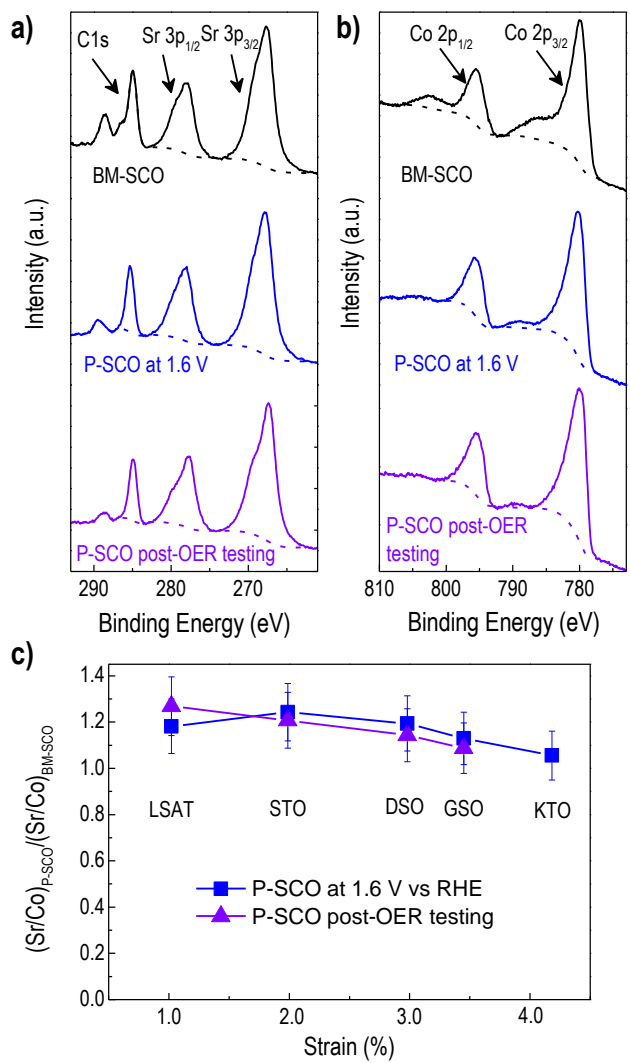

**Figure S5. Evidence of chemical stability throughout OER testing.** Representative XPS scans on an STO substrate of **(a)** Sr 3p and **(b)** Co 2p for BM-SCO, P-SCO oxidized at 1.6 V vs RHE for 5 minutes, and P-SCO after being initially oxidized and tested for OER activity and stability (post-test). Backgrounds (--) are determined using a Shirley fit. **(c)** To verify chemical stability on multiple substrates, the ratio of Sr/Co is quantified for both the BM-SCO and SCO films using the Sr 3p and Co 2p peaks. Similar ratios were also obtained using the Sr 3d peak. By normalizing the Sr/Co ratio of the P-SCO films to BM-SCO, two main points can be determined. First, by having a value greater than 1, there are indications of Sr enrichment near the surface. Second, the normalized ratios for both the initially oxidized P-SCO and post-test P-SCO are within statistical error found using quadrature methods. This indicates that the film is chemically stable near the surface.

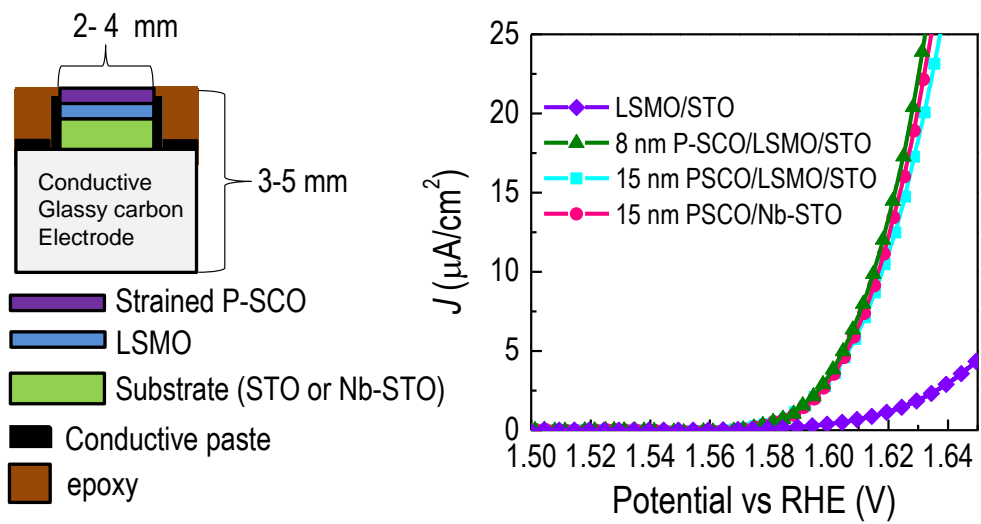

**Figure S6. Testing possible LSMO and conductivity effects of thin film setup on OER activity.** This thin film testing method does not rely on a carbon binder (which can degrade the oxide and affect the OER current) for conductivity. Rather, current passes through the film to a conducting LSMO underlayer grown over a non-conducting substrate, such as STO. To ensure that the LSMO is not affecting activity, it was directly measured by itself. A thinner 8 nm P-SCO film was grown on it and compared to the standard 15 nm used in the study. Finally, a 15 nm P-SCO film was grown directly on a conducting Nb-doped STO substrate without LSMO. The similar results among the films with P-SCO indicate that conductivity is not a likely source of error for OER activity. Additionally, since, as expected, the activity of LSMO is much lower than any of the P-SCO films, if by any chance it were exposed to the solution, it's impact is minimal.

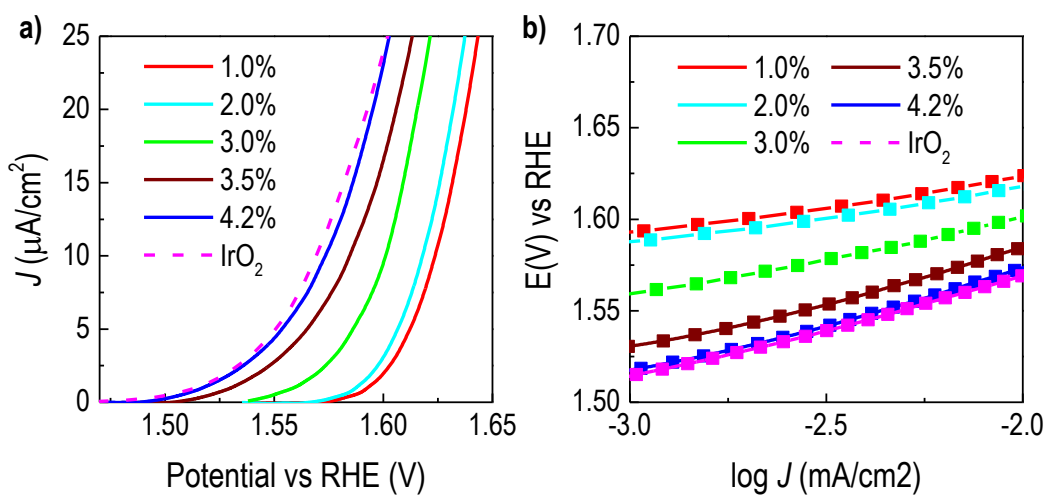

**Figure S7. Polarization and Tafel curves for strained P-SCO and IrO$_2$.** The (a) polarization and (b) Tafel curves for both strained P-SCO and IrO$_2$ are similar. The Tafel slopes for all curves are approximately ~40 mV, suggesting that the rate-determining mechanism is similar for all films, and that an increase in oxygen vacancies increases the exchange current corresponding to this OER reaction. In the interest of space, the curves for $\varepsilon = 0\%$ are not shown, being almost identical to $\varepsilon = 1\%$. Interestingly, the highest strained ($\varepsilon = 4.2\%$), most active P-SCO shows comparable activity to IrO$_2$.